\documentclass[aps,pre,reprint,superscriptaddress,showpacs,amsmath,amssymb,twocolumn,preprintnumbers]{revtex4-1}

\usepackage{graphics,graphicx}

\begin{document}
\preprint{Physical Review E}

\title{Logarithmic scaling for fluctuations of a scalar concentration in wall turbulence}

\author{Hideaki Mouri}
\affiliation{Meteorological Research Institute, Nagamine, Tsukuba 305-0052, Japan}
\affiliation{Graduate School of Science, Kobe University, Rokkodai, Kobe 657-8501, Japan}
\author{Takeshi Morinaga}
\affiliation{Meteorological Research Institute, Nagamine, Tsukuba 305-0052, Japan}
\author{Toshimasa Yagi}
\affiliation{Meteorological Research Institute, Nagamine, Tsukuba 305-0052, Japan}
\author{Kazuyasu Mori}
\affiliation{Meteorological Research Institute, Nagamine, Tsukuba 305-0052, Japan}


\begin{abstract}
Within wall turbulence, there is a sublayer where the mean velocity and the variance of velocity fluctuations vary logarithmically with the height from the wall. This logarithmic scaling is also known for the mean concentration of a passive scalar. By using heat as such a scalar in a laboratory experiment of a turbulent boundary layer, the existence of the logarithmic scaling is shown here for the variance of fluctuations of the scalar concentration. It is reproduced by a model of energy-containing eddies that are attached to the wall.
\end{abstract}

\maketitle

%
%
%
%

\section{Introduction} \label{S1}

This is an experimental study of a behavior of a passive scalar within wall turbulence of an incompressible fluid, especially within a turbulent boundary layer over a flat wall. We set the $x$-$y$ plane at the wall, set the $x$ direction along the mean stream, and use $u(z)$ and $w(z)$ to denote velocity fluctuations in the streamwise and wall-normal directions at a height $z$ from the wall. If the turbulence is stationary and its Reynolds number is high enough, it has a sublayer with some constant $\rho u_{\ast}^2$ for the mean rate of turbulent momentum transfer, i.e., for the momentum flux $\rho \langle -uw \rangle$. Here $\rho$ is the mass density, $u_{\ast}$ is the friction velocity, and $\langle \cdot \rangle$ denotes an average.

Within this constant-flux sublayer, the friction velocity $u_{\ast}$ serves as a characteristic constant in units of velocity, while there is no characteristic constant in units of length. Then, the mean streamwise velocity $U(z)$ obeys a relation $dU/dz \varpropto u_{\ast}/z$ \cite{ll59,s48}. Since any similar relation such as for $dU^2/dz$ is not Galilean invariant \cite{o01}, we exclusively have
\begin{subequations}
\label{eq1}
\begin{equation}
\label{eq1a}
\frac{U(z_1)-U(z_2)}{u_{\ast}} = \frac{1}{\kappa_U} \ln \left( \frac{z_1}{z_2} \right) .
\end{equation}
The von K\'arm\'an constant $\kappa_U$ appears to be universal \cite{mmhs13} because its value of $0.39 \pm 0.02$ is common among various classes of wall turbulence, e.g., pipe flows, channel flows, and boundary layers \cite{mmhs13,hvbs13,sf13,bvhs14,lm15}, and is also common between the cases of smooth and rough walls \cite{hvbs13}.

The same scaling is observed for the mean concentration ${\mit\Theta}(z)$ of a passive scalar if its value at the wall $z=0$ is retained constant, ${\mit{\Theta}}_0 \ne 0$ \cite{my71}. We use $\theta(z)$ to denote fluctuations of the concentration and define the mean rate of turbulent transfer of the scalar as $\langle \theta w(z) \rangle$. Within the constant-flux sublayer, $\langle \theta w(z) \rangle$ does not vary and leads to a characteristic constant $\theta_{\ast} = \langle \theta w \rangle /u_{\ast}$. Since ${\mit\Theta}(z)-{\mit{\Theta}}_0$ obeys a relation $d({\mit\Theta}-{\mit\Theta}_0)/dz = d{\mit\Theta}/dz \varpropto \theta_{\ast}/z$ \cite{ll59}, which is exclusively invariant under any shift of ${\mit\Theta}_0$ \cite{mdce06}, we have
\begin{equation}
\label{eq1b}
\frac{{\mit\Theta}(z_1)-{\mit\Theta}(z_2)}{\theta_{\ast}} = -\frac{1}{\kappa_{\mit\Theta}} \ln \left( \frac{z_1}{z_2} \right).
\end{equation}
\end{subequations}
Between Eqs.~(\ref{eq1a}) and (\ref{eq1b}), the sign is opposite because the direction of the scalar flux $\langle \theta w \rangle$ is defined oppositely to that of the momentum flux $\rho \langle -uw \rangle$. The constant $\kappa_{\mit\Theta}$ appears to be universal and is related to the von K\'arm\'an constant $\kappa_U$ through the turbulent Prandtl number in the neutral-stability limit \cite{ll59}, $\kappa_U / \kappa_{\mit\Theta} \simeq 0.8$ \cite{my71}, although its uncertainty is yet as significant as $0.1$ \cite{zekre13}.

Recently, laboratory experiments and field observations have established another logarithmic scaling for the variance $\langle u^2(z) \rangle$ of fluctuations of the streamwise velocity \cite{mmhs13,hvbs12,hvbs13},
\begin{subequations}
\begin{equation}
\label{eq2a}
\frac{\langle u^2(z_1) \rangle - \langle u^2(z_2) \rangle}{u_{\ast}^2} = -C_{u^2} \ln \left( \frac{z_1}{z_2} \right) .
\end{equation}
The constant $C_{u^2}$ appears to be universal \cite{mmhs13}. From the existing data \cite{mmhs13,hvbs13}, the uncertainty-weighted average is obtained as $C_{u^2} \simeq 1.25 \pm 0.03$.

Thus, a logarithmic scaling is not restricted to averages like $U(z)$ and ${\mit\Theta}(z)$. It is actually observed as well for the variance of pressure fluctuations \cite{jm08}. The scaling is also expected for the variance $\langle \theta^2(z) \rangle$ of fluctuations of the above scalar \cite{m15}, 
\begin{equation}
\label{eq2b}
\frac{\langle \theta^2(z_1) \rangle - \langle \theta^2(z_2) \rangle}{\theta_{\ast}^2} = -C_{\theta^2} \ln \left( \frac{z_1}{z_2} \right) ,
\end{equation}
\end{subequations}
because Eq.~(\ref{eq1b}) for the mean scalar concentration ${\mit\Theta}(z)$ is analogous to Eq.~(\ref{eq1a}) for the mean streamwise velocity $U(z)$.

The above Eq.~(\ref{eq2b}) is confirmed here for a turbulent boundary layer in a wind tunnel. Its floor was heated or cooled slightly so as to use the heat as a passive scalar (Sec.~\ref{S2}). Over this floor, the variance of the air temperature $\langle \theta^2(z) \rangle$ does scale logarithmically with the height $z$ (Sec.~\ref{S3}). To explain any scaling of such a variance, however, there is no exclusive relation like $d{\mit\Theta}/dz \varpropto \theta_{\ast}/z$. In fact, $\langle \theta^2(z) \rangle$ itself is invariant under any shift of ${\mit\Theta}_0$. We instead use a model of energy-containing eddies (Sec.~\ref{S4}), which had predicted the logarithmic scaling of the velocity variance $\langle u^2(z) \rangle$ \cite{t76}. The implication of these scaling laws is also to be remarked in Sec.~\ref{S5}.

\begin{figure}[bp]
\resizebox{8.2cm}{!}{\includegraphics*[4.5cm,5.5cm][16.0cm,27.5cm]{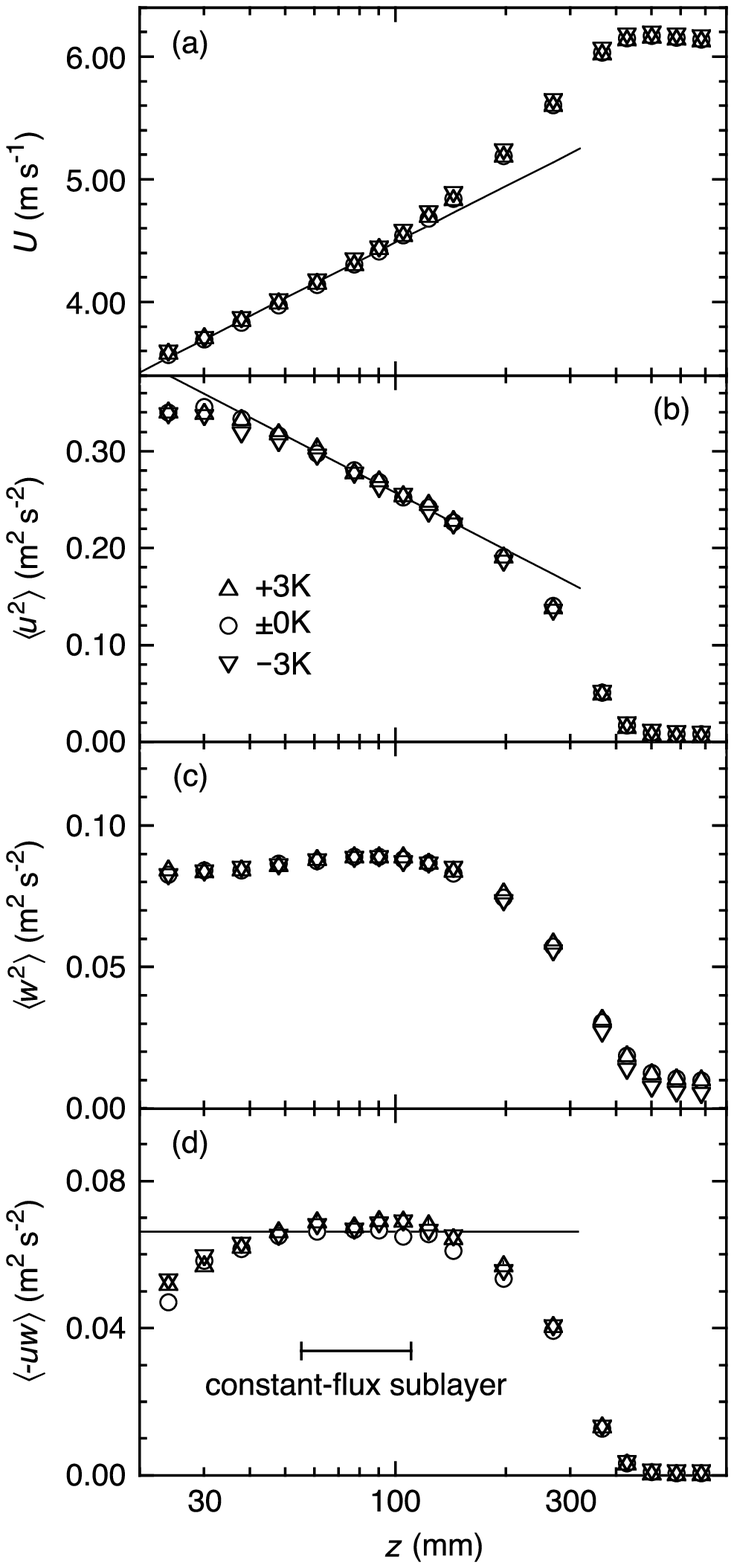}}
\caption{\label{f1} Velocity statistics $U$, $\langle u^2 \rangle$, $\langle w^2 \rangle$, and $\langle -uw \rangle$ as a function of the height $z$. They have been obtained by heating the floor as ${\mit\Theta}_0 - {\mit\Theta}_{\infty} = +3$\,K ($\triangle$), by cooling the floor as $-3$\,K ($\triangledown$), or by neither heating or cooling the floor ($\circ$). For the constant-flux sublayer of the last case, the straight lines denote Eqs.~(\ref{eq1a}) and (\ref{eq2a}) with the parameter values in Table \ref{t1}. The range of this sublayer is also shown. Although error bars of $1\sigma$ lie on all the data, none of them is discernible. }
\end{figure} 

\begin{table}[tbp]
\begingroup
\squeezetable
\caption{\label{t1} Parameter $\nu$ and those of $U(z)/u_{\ast} = (1/\kappa_U) \times \ln (z /z_0)$ for $\kappa_U = 0.39$ with $u_{\ast}^2 = \langle -uw \rangle$, $\langle u^2(z) \rangle /u_{\ast}^2 = B_{u^2} - C_{u^2} \ln (z /\delta_{99})$, and $\langle w^2(z) \rangle /u_{\ast}^2 = B_{w^2}$ within the constant-flux sublayer from $z = 61$\,mm to $105$\,mm, obtained by heating the floor (${\mit\Theta}_0 - {\mit\Theta}_{\infty} = +3$\,K), by cooling the floor ($-3$\,K), or by neither heating or cooling the floor ($\pm 0$\,K).}
\begin{ruledtabular}
\begin{tabular}{ccccc}
                & Unit                       & $+3$\,K           & $\pm 0$\,K        & $-3$\,K           \\ \hline
$\nu$           & cm$^2$\,s$^{-1}$           & $0.150 \pm 0.002$ & $0.145 \pm 0.001$ & $0.155 \pm 0.001$ \\
$u_{\ast}$      & m\,s$^{-1}$                & $0.263 \pm 0.001$ & $0.257 \pm 0.001$ & $0.260 \pm 0.001$ \\
$z_0$           & mm                         & $0.124 \pm 0.004$ & $0.111 \pm 0.001$ & $0.116 \pm 0.003$ \\
$\delta_{99}$   & mm                         & $406   \pm 1$     & $406   \pm 1$     & $404   \pm 1$     \\            
$B_{u^2}$       &                            & $2.03  \pm 0.18$  & $2.09  \pm 0.06$  & $2.17  \pm 0.18$  \\
$C_{u^2}$       &                            & $1.24  \pm 0.11$  & $1.29  \pm 0.05$  & $1.15  \pm 0.11$  \\
$B_{w^2}$       &                            & $1.29  \pm 0.01$  & $1.34  \pm 0.01$  & $1.29  \pm 0.01$  \\
\end{tabular}
\end{ruledtabular}
\endgroup
\end{table}

\section{Experiment} \label{S2}

The experiment was carried out in an open-return wind tunnel of the Meteorological Research Institute. We use coordinates $x$, $y$, and $z$ in the streamwise, spanwise, and floor-normal directions. The origin $x = y = z = 0$\,m is taken on the floor center at the upstream end of the test section of the tunnel. Its size is ${\mit\Delta} x = 18$\,m, ${\mit\Delta} y = 3$\,m, and ${\mit\Delta} z = 2$\,m. The cross section ${\mit\Delta} y \times {\mit\Delta} z$ remains the same upstream to $x = -4$\,m.

On the entire floor from $x = -4$\,m to $x = +18$\,m with an interval of $100$\,mm, aluminium rods oriented to the spanwise direction were set as roughness. The diameter of each of the rods was $3$\,mm.

The boundary layer was formed over that floor. We set the incoming flow velocity to be $6$\,m\,s$^{-1}$ and obtained all the data at a horizontal position where the turbulent boundary layer was well developed, i.e., $x = +14$\,m and $y = 0$\,m.

We measured the streamwise velocity $U+u$ and the floor-normal velocity $w$ over a range of the height $z$ with use of a laser Doppler anemometer (Dantec, model F60 with 60X17). Its sensing volume has the diameter of $120$\,$\mu$m and the length of $1.5$\,mm that was oriented to the spanwise direction.

For the above and other measurements at each height $z$, the sampling rate was set at $100$ or $200$\,Hz. The duration was set at $150$ or $300$\,s. With intervals, they were repeated $3$ or $4$ times. Among the repeated measurements, scattered were the averages, variances, and other cumulants. This scatter is to be used for our error estimations in a standard manner \cite{br03}.

The results are shown by circles in semi-log plots of Fig. \ref{f1}. Although error bars of $1\sigma$ lie on all the circles, none of them is greater than the circle so as to be discernible. We also summarize the values of the flow parameters in Table \ref{t1}. The errors of $1\sigma$ are provided, except for the kinematic viscosity $\nu$ for which we have provided the range of the observed value.

At least between the heights $z \simeq 60$\,mm and $100$\,mm, there is the constant-flux sublayer for which $\langle -uw \rangle$ in Fig.~\ref{f1}(d) is almost constant. The values of $\langle -uw \rangle$ in this sublayer are used to estimate the friction velocity $u_{\ast}$.

The sublayer exhibits a logarithmic scaling for $U(z)$ in Fig.\,\ref{f1}(a) and for $\langle u^2(z) \rangle$ in Fig.\,\ref{f1}(b). As for the case of $U(z)$, the data points fall on the straight line of Eq.~(\ref{eq1a}) if the standard value of $\kappa_U = 0.39$ is adopted from the literature \cite{mmhs13,sf13,bvhs14,lm15,hvbs13}. The accuracy is high, as seen in Table~\ref{t1} where small is the uncertainty of the parameter $z_0$ that has been estimated from $U(z)/u_{\ast} = (1/\kappa_U)\ln (z /z_0)$ \cite{ll59,my71}. To the data points of $\langle u^2(z) \rangle$, we fit a straight line in the form of $\langle u^2(z) \rangle /u_{\ast}^2 = B_{u^2} - C_{u^2} \ln (z /\delta_{99})$ \cite{t76}. Here $\delta_{99}$ is the height at which $U(z)$ is 99\% of its maximum. The resultant estimates of $B_{u^2}$ and $C_{u^2}$ in Table \ref{t1} are consistent with uncertainty-weighted averages of the existing values \cite{mmhs13,hvbs13}, i.e., $B_{u^2} \simeq 2.2 \pm 0.2$ and $C_{u^2} \simeq 1.25 \pm 0.03$. We have ignored the values of $B_{u^2} \simeq 1.5 \pm 0.1$ in pipe flows, which are known to be distinct from those in other classes of wall turbulence \cite{mmhs13,cmmvs15}.

At the lower height $z \lesssim 40$\,mm, the velocity field was affected by the roughness. This was ascertained by measuring $U+u$ and $w$ at a horizontal position slightly shifted in the streamwise direction. However, deviation from the logarithmic scaling is small and slow \cite{mmhs13}. It pretends as if it extends to $z \lesssim 40$\,mm especially in the case of $U(z)$ in Fig.~\ref{f1}(a).

Having confirmed that the constant-flux sublayer was formed at least from $z \simeq 60$\,mm to $100$\,mm, with parameter values consistent with those in the literature \cite{mmhs13,sf13,bvhs14,lm15,hvbs13}, we are to explain our measurements for the scalar.

The wind tunnel is capable of controlling the temperature of the floor of the test section, by circulating heated or cooled liquid through the panels of the floor. We used this capability to retain the floor temperature higher or lower by 3\,K than the temperature of the ambient air, which was monitored at a position upstream of the test section. The ambient air temperature was not constant but varied slowly  by $0.1$\,K per an hour in the most significant case. As a result, we had to adjust repeatedly the floor temperature. It was accordingly fluctuating, albeit with an amplitude as small as $\pm 0.02$\,K for the panel of the floor at around the measurement position of $x = +14$\,m and $y = 0$\,m.

For this setting, we again measured the flow velocities $U+u$ and $w$. The results are shown by triangles in Fig.~\ref{f1}. They are not distinguishable from the circles, i.e., data points for our results obtained without heating or cooling the floor (see also Table \ref{t1}). Thus, since the velocity field was not so affected by the heat, it is almost safe to regard the heat as a passive scalar. The air temperature is to be used as the scalar concentration ${\mit\Theta}+\theta$.

Then, ${\mit\Theta}+\theta$ was measured with use of a cold-wire thermometer (Dantec, model 90C20 with 55P11). Its sensing volume has the diameter of $5$\,$\mu$m and the length of $1.25$\,mm that was oriented to the spanwise direction. The mean rate $H_0$ of heat transfer across the surface of the floor was also measured at $x \simeq +14$\,m and $y \simeq +0.4$\,m. We used two sensors (Etodenki, model M55A), each of which has the size of $50\mbox{\,mm} \times 50 \mbox{\,mm} \times 0.7$\,mm. These measurements were not simultaneous to the velocity measurements, albeit under almost the same conditions, e.g., the temperature of the ambient air ${\mit\Theta}_{\infty}$ in the range of $291 \pm 6$\,K.

\begin{figure}[bp]
\resizebox{8.2cm}{!}{\includegraphics*[4.5cm,5.5cm][16.0cm,27.5cm]{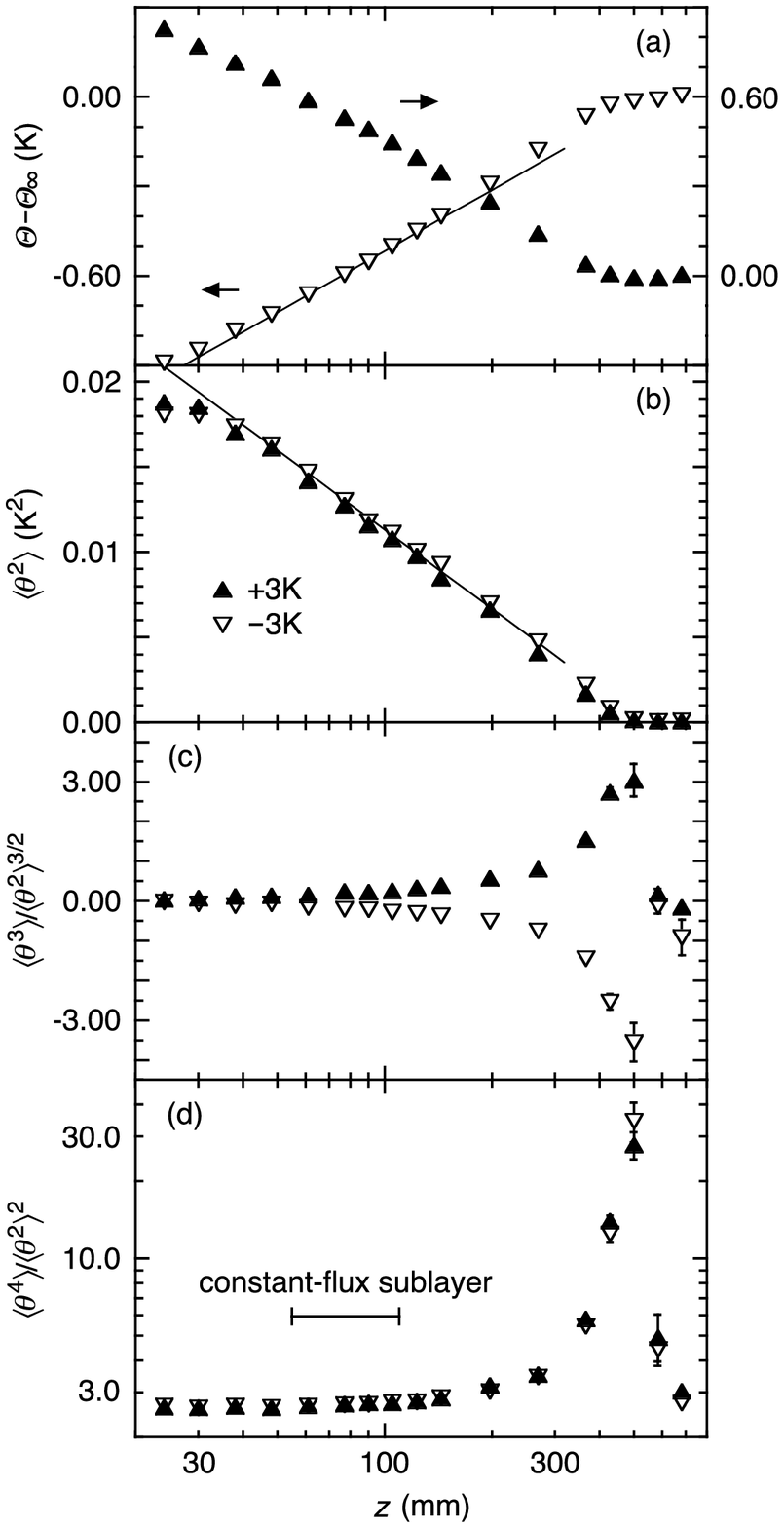}}
\caption{\label{f2} Temperature statistics ${\mit\Theta}-{\mit\Theta}_{\infty}$, $\langle \theta^2 \rangle$, $\langle \theta^3 \rangle /\langle \theta^2 \rangle^{3/2}$, and $\langle \theta^4 \rangle /\langle \theta^2 \rangle^2$ as a function of the height $z$. They have been obtained by heating the floor as ${\mit\Theta}_0 - {\mit\Theta}_{\infty} = +3$\,K ($\blacktriangle$) or by cooling the floor as $-3$\,K ($\triangledown$). For the constant-flux sublayer of the latter case, the straight lines denote Eqs.~(\ref{eq1b}) and (\ref{eq2b}) with the parameter values in Table \ref{t2}. The range of this sublayer is also shown. Although error bars of $1\sigma$ lie on all the data, only a few of them is discernible. }
\end{figure} 

\begin{table}[tbp]
\begingroup
\squeezetable
\caption{\label{t2} Parameters of ${\mit\Theta}(z)/\theta_{\ast} - {\mit\Theta}_{\infty}/\theta_{\ast} = B_{\mit\Theta} - (1/\kappa_{\mit\Theta}) \times \ln (z /\delta_{99})$ and $\langle \theta^2(z) \rangle /\theta_{\ast}^2 = B_{\theta^2} - C_{\theta^2} \ln (z /\delta_{99})$ within the constant-flux sublayer from $z = 61$\,mm to $105$\,mm, obtained by heating the floor (${\mit\Theta}_0 - {\mit\Theta}_{\infty} = +3$\,K) or by cooling the floor ($-3$\,K), where the value of $\delta_{99}$ is from Table \ref{t1} for the case of neither heating or cooling the floor.}
\begin{ruledtabular}
\begin{tabular}{cccc}
                                       & Unit   & $+3$\,K             & $-3$\,K  \\ \hline
$\theta_{\ast}/\kappa_{\mit\Theta}$    & K      & $+0.260 \pm 0.015$  & $-0.295 \pm 0.010$ \\
$\kappa_{\mit\Theta}   B_{\mit\Theta}$ &        & $0.386  \pm 0.124$  & $0.351  \pm 0.068$  \\
$\kappa_{\mit\Theta}^2 B_{\theta^2}$   &        & $0.0335 \pm 0.0063$ & $0.0223 \pm 0.0091$ \\
$\kappa_{\mit\Theta}^2 C_{\theta^2}$   &        & $0.0934 \pm 0.0105$ & $0.0770 \pm 0.0080$
\end{tabular}
\end{ruledtabular}
\endgroup
\end{table}

\section{Results} \label{S3}

The results of our measurements of the air temperature are shown by triangles in Fig.~\ref{f2}. For the constant-flux sublayer from $z \simeq 60$\,mm to $100$\,mm, we estimate the values of the parameters of Eqs.~(\ref{eq1b}) and (\ref{eq2b}). They are shown by straight lines for the case of heating the tunnel floor. We also summarize the parameter values in Table \ref{t2}, where $\theta_{\ast}/\kappa_{\mit\Theta}$ and $\kappa_{\mit\Theta}^2 C_{\theta^2}$ are provided instead of $\theta_{\ast}$ and $C_{\theta^2}$ because the value of $\kappa_{\mit\Theta}$ is not yet so certain \cite{zekre13} as has been remarked in Sec.~\ref{S1}.

Figure \ref{f2}(a) is a semi-log plot for the mean temperature ${\mit\Theta}(z)$ with respect to the temperature of the ambient air ${\mit\Theta}_{\infty}$, which was less fluctuating than the floor temperature ${\mit\Theta}_0$ (Sec.~\ref{S2}). The constant-flux sublayer does exhibit the logarithmic scaling of Eq.~(\ref{eq1b}).

To consider the parameter $\theta_{\ast}/\kappa_{\mit\Theta}$ of Eq.~(\ref{eq1b}), we use the mean rate of heat transfer measured across the floor surface, $H_0 = \pm 48$\,W\,m$^{-2}$. Under the pressure of $1$\,atm, the air at $290$\,K has the mass density $\rho = 1.2$\,kg\,m$^{-3}$ and the isobaric specific heat $c_p = 1.0 \times 10^3$\,J\,kg$^{-1}$\,K$^{-1}$. By also using the friction velocity $u_{\ast} = 0.26$\,m\,s$^{-1}$ (Table~\ref{t1}), we obtain $\pm 0.15$\,K as the values of $\theta_{\ast} = H_0/ (c_p \rho u_{\ast})$ \cite{ll59,my71}. They are consistent with our estimates of $\theta_{\ast}/\kappa_{\mit\Theta}$ in Table \ref{t2} because $\kappa_{\mit\Theta} \simeq 0.5 \pm 0.1$ is implied from the estimates of $\kappa_U / \kappa _{\mit\Theta} \simeq 0.8 \pm 0.1$ in the literature  \cite{my71,zekre13}.

Figure \ref{f2}(b) is our main result, i.e., a semi-log plot for the variance $\langle \theta^2(z) \rangle$ of the temperature fluctuations. The variance decreases logarithmically with an increase in the height $z$. Those for ${\mit\Theta}_0 - {\mit\Theta}_{\infty} = +3$\,K and $-3$\,K are almost indistinguishable. We have thus confirmed the logarithmic scaling of $\langle \theta^2(z) \rangle$ within the constant-flux sublayer in accordance with Eq.~(\ref{eq2b}).

The logarithmic scaling of $\langle \theta^2(z) \rangle$ appears in Fig.~\ref{f2}(b) to extend beyond the height range of the constant-flux sublayer that has been determined from the measurement of $\langle -uw(z) \rangle$. This is also true for the logarithmic scaling of ${\mit\Theta}(z)$ in Fig.~\ref{f2}(a). From these scalings, deviation might be small and slow as has been observed for $U(z)$ in Sec.~\ref{S2} \cite{mmhs13}. There is yet a possibility that $\langle \theta w(z)\rangle / \langle -uw(z) \rangle^{1/2}$ is somehow retained close to $\theta_{\ast}$ throughout those heights $z$. It would be of interest to measure the scalar flux $\langle \theta w(z) \rangle$, albeit difficult if a high accuracy is necessary.

The value estimated for the parameter $\kappa_{\mit\Theta}^2 C_{\theta^2}$ of Eq. (\ref{eq2b}) is not exactly equal between the cases of ${\mit\Theta}_0 - {\mit\Theta}_{\infty} = +3$\,K and $-3$\,K (Table \ref{t2}) because the heat does not serve exactly as a passive scalar. By interpolating these two cases to the limit $\theta_{\ast} \rightarrow 0$, the exact value is estimated to be $\kappa_{\mit\Theta}^2 C_{\theta^2} \simeq 0.086 \pm 0.007$.

Lastly, the skewness $\langle \theta^3(z) \rangle / \langle \theta^2(z) \rangle^{3/2}$ and the flatness $\langle \theta^4(z) \rangle / \langle \theta^2(z) \rangle^2$ of the temperature fluctuations are shown in Figs.~2(c) and 2(d). Within the constant-flux sublayer, they are respectively close to the Gaussian values of $0$ and $3$ \cite{ad77}, implying that the distribution of $\theta$ is closely Gaussian. They are enhanced at around the edge of the boundary layer \cite{ad77}, the height of which varies significantly in space and in time. Similar results are known for the skewness and flatness of the velocity fluctuations \cite{ff96,mthk03}.

\begin{figure}[bp]
\resizebox{8.cm}{!}{\includegraphics*[3.9cm,20.8cm][16.5cm,26.3cm]{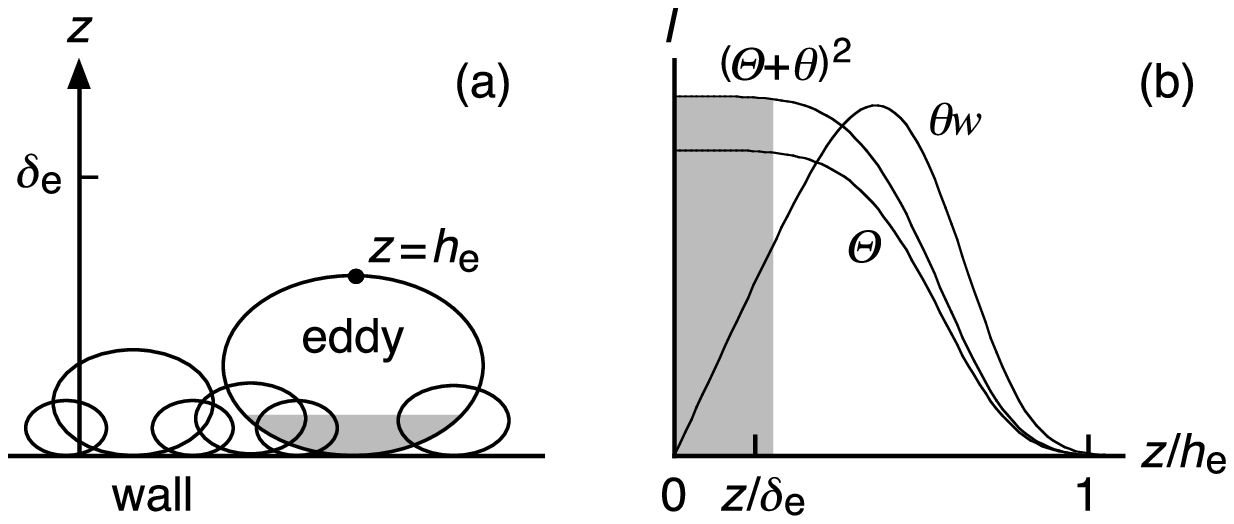}}
\caption{\label{f3} Schematics of attached eddies and of the moments $I_{\mit\Theta}(z/h_{\rm e})$, $I_{({\mit\Theta}+\theta)^2}(z/h_{\rm e})$, and $I_{\theta w}(z/h_{\rm e})$ defined in Eqs.~(\ref{eq5b}), (\ref{eq8b}), and (\ref{eq10b}), where grey areas correspond to the undermost portions of the eddies. }
\end{figure} 

\section{Discussion} \label{S4}

Having confirmed the logarithmic scaling of $\langle \theta^2(z) \rangle$ in Eq.~(\ref{eq2b}), we are to reproduce this scaling by using the attached-eddy hypothesis in its original form \cite{t76}. It had predicted not only the logarithmic scaling of $\langle u^2(z) \rangle$ in Eq.~(\ref{eq2a}) but also $\langle w^2(z) \rangle \varpropto u_{\ast}^2$, which is actually observed in Fig.~\ref{f1}(c).

This hypothesis is a model of a random superposition of energy-containing eddies that are attached to the wall as illustrated in Fig.~\ref{f3}(a). While their velocity fields have an identical shape with a common characteristic velocity $u_{\ast}$, their sizes are distributed with no characteristic size. For example, if $\mbox{\boldmath{$x$}}_{\rm e} = (x_{\rm e}, y_{\rm e}, h_{\rm e})$ lies at the highest position of an eddy, its wall-normal velocity $w_{\rm e}$ is given by a function $f_w$ for any position $\mbox{\boldmath{$x$}} = (x, y, z)$ as
\begin{subequations}
\label{eq3}
\begin{equation}
\label{eq3a}
\frac{w_{\rm e}(\mbox{\boldmath{$x$}})}{u_{\ast}} = f_w \left( \frac{\mbox{\boldmath{$x$}} - \mbox{\boldmath{$x$}}_{\rm e}}{h_{\rm e}} \right).
\end{equation}
The height $h_{\rm e}$ of this eddy is used as its size. Since $w_{\rm e}$ is blocked by the wall, we require $f_w = 0$ at $z = 0$ so that $f_w \varpropto z/h_{\rm e}$ at $z/h_{\rm e} \ll 1$.

The scalar field ${\mit\Theta}_{\rm e} + \theta_{\rm e}$ of each eddy is also assumed to have an identical shape with a common characteristic concentration $\theta_{\ast}$,
\begin{equation}
\label{eq3b}
\frac{{\mit\Theta}_{\rm e}(\mbox{\boldmath{$x$}}) + \theta_{\rm e}(\mbox{\boldmath{$x$}})}{\theta_{\ast}} 
= f_{{\mit\Theta}+\theta} \left( \frac{\mbox{\boldmath{$x$}} - \mbox{\boldmath{$x$}}_{\rm e}}{h_{\rm e}} \right).
\end{equation}
\end{subequations}
We have included ${\mit\Theta}_{\rm e}(\mbox{\boldmath{$x$}})$ so as to consider the mean concentration ${\mit\Theta}(z)$. Since its value at the wall $z = 0$ has to be constant, i.e., ${\mit{\Theta}}_0 \ne 0$ (Sec.~\ref{S1}), the function $f_{{\mit\Theta}+\theta}$ does not depend on $z/h_{\rm e}$ but depends only on $(x-x_{\rm e})/h_{\rm e}$ and $(y - y_{\rm e})/h_{\rm e}$ at $z/h_{\rm e} \ll 1$.

The eddy size is distributed from $h_{\rm e} \rightarrow 0$ to $h_{\rm e} = \delta_{\rm e}$. Here $\delta_{\rm e}$ is the thickness of the wall turbulence, e.g., $\delta_{99}$ in case of a boundary layer. On the wall, the distribution of the eddies is random and independent. They are allowed to overlap one another because they do not have to be coherent.

From the random and independent distribution of the eddies, it follows that the entire scalar field is a superposition of those of the individual eddies. The asymptotic laws for $z/\delta_{\rm e} \rightarrow 0$ are to be regarded as the laws for the constant-flux sublayer \cite{t76}, by ignoring the roughness or viscosity close to the wall.

The mean concentration ${\mit\Theta}(z)$ is written as an integration from $h_{\rm e} = z$ to $h_{\rm e} = \delta_{\rm e}$,
\begin{equation}
\label{eq4}
\frac{{\mit\Theta}(z)}{\theta_{\ast}} 
= \! \int^{\delta_{\rm e}}_z \! \! \! dh_{\rm e} 
     \left[ n_{\rm e}(h_{\rm e}) \! \! \iint \! \! dx_{\rm e} \, dy_{\rm e} \,  f_{{\mit\Theta}+\theta} \! \left( \frac{\mbox{\boldmath{$x$}} - \mbox{\boldmath{$x$}}_{\rm e}}{h_{\rm e}} \right) \right].
\end{equation}
Here $n_{\rm e}(h_{\rm e})$ is the number density of eddies of size $h_{\rm e}$ per unit area of the wall. With use of a constant $N_{\rm e}$, we have $n_{\rm e}(h_{\rm e}) = N_{\rm e} h_{\rm e}^{-3}$ \cite{t76}. This is because, apart from $h_{\rm e}$ and $N_{\rm e}$, the constant-flux sublayer has no quantity to affect the value of the number density $n_{\rm e}(h_{\rm e})$. Then,
\begin{subequations}
\label{eq5}
\begin{equation}
\label{eq5a}
\frac{{\mit\Theta}(z)}{\theta_{\ast}} = N_{\rm e} \! \int^{\delta_{\rm e}}_z \! \frac{dh_{\rm e}}{h_{\rm e}} \, I_{\mit\Theta} \! \left( \frac{z}{h_{\rm e}} \right),
\end{equation}
with the contribution from eddies of size $h_{\rm e}$,
\begin{equation}
\label{eq5b}
I_{\mit\Theta} \! \left( \frac{z}{h_{\rm e}} \right) 
\! = \!
\iint \! \frac{dx_{\rm e}}{h_{\rm e}} \, \frac{dy_{\rm e}}{h_{\rm e}} \, f_{{\mit\Theta}+\theta} \! \left( \frac{\mbox{\boldmath{$x$}} - \mbox{\boldmath{$x$}}_{\rm e}}{h_{\rm e}} \right) .
\end{equation}
\end{subequations}
By using $\zeta = z/h_{\rm e}$ and hence $d \zeta / \zeta = -dh_{\rm e}/h_{\rm e}$, we rewrite Eq.~(\ref{eq5a}) as
\begin{equation}
\label{eq6}
\frac{{\mit\Theta}(z)}{\theta_{\ast}} = N_{\rm e} \! \int^1_{z/\delta_{\rm e}} \! \frac{d\zeta}{\zeta} \, I_{\mit\Theta} (\zeta ).
\end{equation}
The condition at $z = 0$ on $f_{{\mit\Theta}+\theta}$ implies $I_{\mit\Theta} (\zeta ) \simeq I_{\mit\Theta}(0) \ne 0$ at $\zeta = z/h_{\rm e} \ll 1$ as illustrated in Fig.~\ref{f3}(b). In the limit $z/\delta_{\rm e} \rightarrow 0$, there is some constant $b_{\mit\Theta}$ such that ${\mit\Theta}(z) / \theta_{\ast} \rightarrow N_{\rm e} [b_{\mit\Theta} - I_{\mit\Theta}(0) \ln (z/ \delta_{\rm e})]$. If we use $B_{\mit\Theta} = N_{\rm e} b_{\mit\Theta}$ and $C_{\mit\Theta} = 1/ \kappa_{\mit\Theta} = N_{\rm e} I_{\mit\Theta}(0)$,
\begin{equation}
\label{eq7}
\frac{{\mit\Theta}(z)}{\theta_{\ast}} \rightarrow B_{\mit\Theta} - \frac{1}{\kappa_{\mit\Theta}} \ln \left( \frac{z}{\delta_{\rm e}} \right)
\quad \mbox{as} \quad
\frac{z}{\delta_{\rm e}} \rightarrow 0.
\end{equation}
This asymptotic relation corresponds to the logarithmic scaling of ${\mit\Theta}(z)$ in Eq.~(\ref{eq1b}).

The variance $\langle \theta^2(z) \rangle$ of fluctuations of the scalar concentration is likewise written as
\begin{subequations}
\label{eq8}
\begin{equation}
\label{eq8a}
\frac{\langle \theta^2(z) \rangle}{\theta^2_{\ast}} 
= N_{\rm e} \! \int^{\delta_{\rm e}}_z \! \frac{dh_{\rm e}}{h_{\rm e}} \left[ I_{({\mit\Theta}+\theta)^2} \! \left( \frac{z}{h_{\rm e}} \right) - I^2_{\mit\Theta} \! \left( \frac{z}{h_{\rm e}} \right) \right],
\end{equation}
with the contribution from eddies of size $h_{\rm e}$ that is made up of $I_{\mit\Theta}(z/h_{\rm e})$ in Eq.~(\ref{eq5b}) and also of
\begin{equation}
\label{eq8b}
I_{({\mit\Theta}+\theta)^2} \! \left( \frac{z}{h_{\rm e}} \right) 
\! = \!
\iint \! \frac{dx_{\rm e}}{h_{\rm e}} \, \frac{dy_{\rm e}}{h_{\rm e}} \, f^2_{{\mit\Theta}+\theta} \! \left( \frac{\mbox{\boldmath{$x$}} - \mbox{\boldmath{$x$}}_{\rm e}}{h_{\rm e}} \right) .
\end{equation}
\end{subequations}
The condition at $z = 0$ on $f_{{\mit\Theta}+\theta}$ implies $I_{({\mit\Theta}+\theta)^2}(\zeta) - I^2_{\mit\Theta}(\zeta) \simeq I_{({\mit\Theta}+\theta)^2}(0) - I^2_{\mit\Theta}(0) > 0$ at $\zeta = z/h_{\rm e} \ll 1$ (Fig.~\ref{f3}). Hence, through a relation similar to Eq.~(\ref{eq6}),
\begin{equation}
\label{eq9}
\frac{\langle \theta^2(z) \rangle}{\theta^2_{\ast}} \rightarrow B_{\theta^2} - C_{\theta^2} \ln \left( \frac{z}{\delta_{\rm e}} \right)
\quad \mbox{as} \quad
\frac{z}{\delta_{\rm e}} \rightarrow 0.
\end{equation}
This asymptotic relation corresponds to the logarithmic scaling of $\langle \theta^2(z) \rangle$ in Eq.~(\ref{eq2b}).

Lastly, we consider the scalar flux $\langle \theta w(z) \rangle$. It is written as
\begin{subequations}
\label{eq10}
\begin{equation}
\label{eq10a}
\frac{\langle \theta w(z) \rangle}{\theta_{\ast} u_{\ast}} 
= N_{\rm e} \! \int^{\delta_{\rm e}}_z \! \frac{dh_{\rm e}}{h_{\rm e}} \, I_{\theta w} \! \left( \frac{z}{h_{\rm e}} \right) ,
\end{equation}
with the contribution from eddies of size $h_{\rm e}$,
\begin{equation}
\label{eq10b}
I_{\theta w} \! \left( \frac{z}{h_{\rm e}} \right) 
\! = \!
\iint \! \frac{dx_{\rm e}}{h_{\rm e}} \, \frac{dy_{\rm e}}{h_{\rm e}} \, f_{{\mit\Theta}+\theta} \! \left( \frac{\mbox{\boldmath{$x$}} - \mbox{\boldmath{$x$}}_{\rm e}}{h_{\rm e}} \right) 
                                                                           f_w                     \! \left( \frac{\mbox{\boldmath{$x$}} - \mbox{\boldmath{$x$}}_{\rm e}}{h_{\rm e}} \right).
\end{equation}
\end{subequations}
The conditions at $z = 0$ on $f_{{\mit\Theta}+\theta}$ and $f_w$ imply $I_{\theta w}(\zeta) \varpropto \zeta$ at $\zeta = z/h_{\rm e} \ll 1$ (Fig.~\ref{f3}). Hence, through a relation similar to Eq.~(\ref{eq6}),
\begin{equation}
\label{eq11}
\frac{\langle \theta w(z) \rangle}{\theta_{\ast} u_{\ast}} \rightarrow B_{\theta w}
\quad \mbox{as} \quad
\frac{z}{\delta_{\rm e}} \rightarrow 0.
\end{equation}
Especially if $B_{\theta w}$ is equal to unity, this asymptotic relation reproduces the constant flux $\langle \theta w \rangle = \theta_{\ast} u_{\ast}$ for the scalar transfer.

While $B_{\mit\Theta}$, $B_{\theta^2}$, and $B_{\theta w}$ are due to eddies of sizes $h_{\rm e}$ comparable to the height $z$, $C_{\mit\Theta} = 1/\kappa_{\mit\Theta}$ and $C_{\theta^2}$ are due to eddies of sizes $h_{\rm e}$ from the height $z$ to the thickness $\delta_{\rm e}$ of the wall turbulence. The integration over such eddies has led to the logarithmic factor, $\ln (z/\delta_{\rm e})$.

Since $C_{\mit\Theta}$ and $C_{\theta^2}$ are determined only by the undermost portions of those eddies (grey areas in Fig.~\ref{f3}), they would be insensitive to the class of the flow configuration for the production of the wall turbulence. This is actually true in the case of $C_{u^2}$ for the velocity variance in Sec.~\ref{S1} \cite{mmhs13}. On the other hand, $B_{\mit\Theta}$ and $B_{\theta^2}$ are affected also by the upper portions of the eddies. They might be sensitive to the class of the flow configuration as in the case of $B_{u^2}$ in Sec.~\ref{S2} \cite{mmhs13,cmmvs15}.

To obtain some constant value for $C_{\theta^2}$, the amplitude $|f_{{\mit\Theta}+\theta}|$ of the function $f_{{\mit\Theta}+\theta}$ has to be $\varpropto 1/N_{\rm e}^{1/2}$ in Eq.~(\ref{eq8}). We thereby obtain $C_{\mit\Theta} = 1/\kappa_{\mit\Theta} = N_{\rm e} I_{\mit\Theta}(0) \varpropto N_{\rm e} |f_{{\mit\Theta}+\theta}| \varpropto N_{\rm e}^{1/2}$ in Eq.~(\ref{eq5}). For the constant $B_{\theta w} = 1$, we require $|f_{{\mit\Theta}+\theta}| |f_w| \varpropto 1/N_{\rm e}$ and hence $|f_w| \varpropto |f_{{\mit\Theta}+\theta}| \varpropto 1/N_{\rm e}^{1/2}$ in Eq.~(\ref{eq10}).

The value of $N_{\rm e}$ is large but remains finite. In Figs.~\ref{f2}(c) and \ref{f2}(d), the skewness and flatness of the fluctuations $\theta$ are close to but are not equal to their Gaussian values \cite{ad77}. The central limit theorem \cite{my71} implies that the fluctuations $\theta$ tend Gaussian as $N_{\rm e}$ tends infinity.

Thus, the logarithmic scaling of ${\mit{\Theta}}(z)$ in Eq.~(\ref{eq1b}), the logarithmic scaling of $\langle \theta^2 (z) \rangle$ in Eq.~(\ref{eq2b}), and the constant value of $\langle \theta w(z) \rangle$ are all explainable by a superposition of attached eddies. Their scalar fields are given by Eq.~(\ref{eq3b}). Along with the power-law distribution of the eddy size $n_{\rm e}(h_{\rm e}) \varpropto h_{\rm e}^{-3}$, the form of Eq.~(\ref{eq3b}) is in accordance with the existence of the characteristic concentration $\theta_{\ast}$ and with the nonexistence of any characteristic constant in units of length. These two are the necessary and sufficient conditions for the logarithmic scaling of the average ${\mit{\Theta}}(z)$ \cite{mdce06}. Since they are not sufficient for the variance $\langle \theta^2(z) \rangle$ (see Sec.~\ref{S1}), the distribution of the eddies on the wall has been set random and independent \cite{t76}. Over such eddies, also to be considered in Sec.~\ref{S5}, cumulants like the variance $I_{({\mit\Theta}+\theta)^2} - I^2_{\mit\Theta}$ are exclusively allowed to be integrated as in Eq.~(\ref{eq8a}). The average $I_{\mit\Theta}$ in Eq.~(\ref{eq5a}) and the covariance $I_{\theta w}$ in Eq.~(\ref{eq10a}) are other examples of the cumulants.

\section{Concluding Remarks} \label{S5}

By using heat as a passive scalar in a laboratory experiment of a turbulent boundary layer, it has been shown that the constant-flux sublayer exhibits the logarithmic scaling of Eq.~(\ref{eq2b}) for the variance $\langle \theta^2 (z) \rangle$ of fluctuations of the scalar concentration. The parameter $\kappa_{\mit\Theta}^2 C_{\theta^2}$ lies at $0.086 \pm 0.007$. We have explained this scaling with use of the attached-eddy hypothesis \cite{t76}, i.e., a model of a random superposition of energy-containing eddies that are attached to the wall.

The attached-eddy hypothesis is applicable not only to boundary layers but to any other class of wall turbulence, e.g., pipe and channel flows. Also within such flows, the logarithmic scaling of $\langle \theta^2 (z) \rangle$ is likely to exist with the same value of $C_{\theta^2}$.

From the attached-eddy hypothesis, a logarithmic scaling is also expected for the higher order cumulants such as $\langle \theta(z)^4 \rangle - 3 \langle \theta(z)^2 \rangle^2$. This is because any cumulant of a sum of random variables is equal to the sum of cumulants of the variables if they are independent of one another \cite{my71}. The $m$th cumulant is equal to some $m$th-order homogeneous polynomial of the first $m$ moments. If we define the moments $I_{({\mit\Theta}+\theta)^n}(z/h_{\rm e})$ at $n = 3, ..., \, m$ as at $n=2$ in Eq.~(\ref{eq8b}), make up the polynomial, e.g., $I_{({\mit\Theta}+\theta)^4} -4I_{({\mit\Theta}+\theta)^3}I_{\mit\Theta} -3I^2_{({\mit\Theta}+\theta)^2} +12I_{({\mit\Theta}+\theta)^2}I^2_{\mit\Theta} -6I^4_{\mit\Theta}$, and integrate it from $h_{\rm e} = z$ to $h_{\rm e} = \delta_{\rm e}$ as in Eq.~(\ref{eq8a}), the resultant cumulant has the logarithmic factor $\ln (z/\delta_{\rm e})$. However, since $|f_{{\mit\Theta}+\theta}| \varpropto 1/N_{\rm e}^{1/2}$ implies $I_{({\mit\Theta}+\theta)^n} \varpropto 1/N_{\rm e}^{n/2}$ (Sec. \ref{S4}), the $m$th cumulant is of the order of $1/N_{\rm e}^{m/2-1}$. It is increasingly negligible with an increase in the order $m$. We could ignore all the cumulants at $m \ge 3$ to assume that the fluctuations $\theta$ are Gaussian as could be assumed for the case of the velocity fluctuations $u$ \cite{m15,mm13}.

The behavior of a passive scalar is generally dissimilar from the corresponding behavior of the velocity field \cite{gfm77,k94}. Nevertheless, as for energy-containing eddies of wall turbulence, our condition ${\mit\Theta}(0) = {\mit{\Theta}}_0$ is analogous to the condition on the mean streamwise velocity, $U(0) = 0$ \cite{aak09}. This leads to some similarities, e.g., the logarithmic scalings of ${\mit\Theta}(z)$ and $U(z)$ \cite{ll59,my71}. Another example would be made up from the logarithmic scalings of $\langle \theta^2(z) \rangle$ and $\langle u^2(z) \rangle$ studied here.

Besides the attached-eddy hypothesis, the logarithmic scaling of the velocity variance $\langle u^2(z) \rangle$ has been explained by some other models \cite{m15,pa77,h12}. Each of them is based on a particular assumption, in addition to the existence of the characteristic velocity $u_{\ast}$ and to the nonexistence of any characteristic constant in units of length. If $u$ and $u_{\ast}$ were replaced respectively with $\theta$ and $\theta_{\ast}$, such models could reproduce the logarithmic scaling of $\langle \theta^2(z) \rangle$.

To a boundary layer, especially to that over a horizontal wall, there is an application. While we have used heat as a passive scalar, the heat would become active if the wall were heated or cooled still more. The boundary layer would become unstable or stable and would have the Monin-Obukhov length $L_{\ast} = - u_{\ast}^3 / [\kappa (g/{\mit\Theta}_0) (H_0/c_p \rho)]$ as some constant \cite{my71}, where $g$ is the gravitational acceleration. As for such a constant-flux sublayer, it has been considered that any scaling is a function of $z/L_{\ast}$. The temperature variance $\langle \theta^2(z) \rangle$ has been predicted to be $\varpropto \theta_{\ast}^2 (-z/L_{\ast})^{-2/3}$ in the unstable limit $z/L_{\ast} \rightarrow -\infty$ \cite{p54} and to be $\varpropto \theta_{\ast}^2$ in the stable limit $z/L_{\ast} \rightarrow +\infty$ \cite{my71}. These laws are not inconsistent with the observations of the atmospheric boundary layer \cite{wci71}. However, they have to be related continuously to the logarithmic scaling obtained here. In addition, at least in a class of free convection where $L_{\ast}$ is exactly equal to $0$, again logarithmic are the mean temperature ${\mit\Theta}(z)$ \cite{abfghlsv12,abx14} and also the temperature variance $\langle \theta^2(z) \rangle$ \cite{hgba14}. It would be of interest to reconsider the scaling of the temperature fluctuations $\theta (z)$ for these unstable and stable boundary layers.

\begin{acknowledgments}
This work was supported in part by KAKENHI Grant No. 17K00526. 
\end{acknowledgments}

\end{document}